\documentclass{PoS}
\pdfoutput=1
\usepackage[sort&compress,numbers,merge]{natbib}
\usepackage{natbib,natbibspacing}
\usepackage{amsmath,amssymb,upgreek}
\usepackage{graphicx}

\newcommand{\Tr}{\operatorname{Tr}}
\renewcommand{\Re}{\mathfrak{Re}}
\newcommand{\vev}[1]{\ensuremath{\left\langle #1 \right\rangle}}
\newcommand{\lvar}{\ensuremath{\underline{\boldsymbol{\updelta}}}}

\title{The Dyson-Schwinger equation of a link variable in lattice Landau gauge theory}

\ShortTitle{The Dyson-Schwinger equation of a link variable in lattice Landau gauge theory}

\author{\speaker{Andr\'e Sternbeck}%
       \thanks{Former address: Institut f\"ur Theoretische Physik, 
               Universit\"at Regensburg, 93040 Regensburg, Germany.}\\
        Theoretisch-Physikalisches Institut, Friedrich-Schiller-Universit\"at Jena, 07743 Jena, Germany\\
        E-mail: \email{andre.sternbeck@uni-jena.de}}

\author{Martin Schaden\\
        Department of Physics, Rutgers University, Newark NJ 07102, USA\\
        E-mail: \email{mschaden@rutgers.edu}}

\author{Valentin Mader\\
        Institut f\"ur Physik, Karl-Franzens-Universit\"at Graz, A-8010 Graz, Austria}
      
\abstract{
We derive the Dyson-Schwinger equation of a link variable in $SU(n)$ lattice gauge theory 
in minimal Landau gauge and confront it with Monte-Carlo data for the different terms. Preliminary results for the 
lattice analog of the Kugo-Ojima confinement criterion is also shown.
}

\FullConference{The 32nd International Symposium on Lattice Field Theory,\\
 23-28 June, 2014\\
 Columbia University New York, NY}

\begin{document}

\section{Introduction}

The Dyson-Schwinger equation (DSE) of a gauge boson was recently revisited \cite{Schaden:2013ffa} in the context of 
the Kugo-Ojima (KO) confinement criterion \cite{Kugo:1979gm,*Kugo:1995km}. It was found that the term which saturates 
the rhs.\ of the DSE for $p\to0$ depends on the phase of a gauge theory \cite{Schaden:2013ffa}. In the Higgs phase, for 
example, physical states saturate the transverse DSE, while in the confining phase only unphysical degrees 
of freedom contribute to the saturation of the DSE. Interestingly, this corollary to the KO criterion is not only true 
for linear covariant gauges, but can also be applied to other gauges and models like maximal Abelian gauge, 
non-covariant Coulomb gauge and in the Gribov-Zwanziger theory \cite{Schaden:2013ffa}.

In an earlier attempt \cite{Sternbeck:2006rd,*Sternbeck:2006cg} we have tried to verify the KO criterion directly on 
the lattice. There however we could not confirm the desired result for the KO function: $u(p)\to -1$ for $p\to0$. Our 
data was more in favor of $u(p)$ reaching a limit somewhere between $-0.6$ and $-0.8$. In the light of 
\cite{Schaden:2013ffa} it was thus natural to revisit this earlier calculation and to derive the exact 
DSE for a link variable in Landau gauge on the lattice. With this one could check if $u(p)$ also on the lattice 
saturates the DSE in the infrared limit. Here we report on our first findings of a still on-going investigation. 

\section{Dyson-Schwinger equation of a lattice link variable in Landau gauge}

For the following it is advantageous to use a notation in which each link 
variable is assigned its defining sites $x$ and $y=x\pm\hat{\mu}$, i.e., we use the notation $U_{xy}\in SU(n)$ rather 
than the usual $U_{x\mu}$. The Wilson gauge action for an $SU(n)$ lattice gauge theory then reads
\begin{equation}
 \label{eq:Wilson}
  S_W[U]=\frac{1}{4g^2_0}\sum^N_{i,j,k,l} P_{ijkl} \quad\text{with}\quad 
   P_{ijkl}=\Tr \left(U_{ij} U_{jk} U_{kl} U_{li}\right)
\end{equation}
where $P$ is a plaquette variable and $g^2_0=2n/\beta$. Since we are interested in Landau gauge we will consider 
links which minimize the (real-valued) Morse potential
\begin{equation}
\label{eq:Morse}
V[U]=-\frac{1}{2}\sum^N_{i,j} \Tr U_{ij}\,.
\end{equation}
These minima fulfill the Landau gauge condition (we use anti-hermitian generators $t^a$ of $SU(n)$)
\begin{equation}
\label{eq:landau}
 0 \stackrel{!}{=} f^a_i=\sum_j \Re\Tr \left(t^a U_{ij}\right)\qquad \forall i,a\,.
\end{equation}
Starting with a thermalized (non-gauged) configuration $U$, gauge-fixing on the lattice is commonly performed 
by an iterative procedure which consecutively gauge-transforms $U\to U^g_{ij}= g_i U_{ij} g^{\dagger}_j$ until 
Eq.~\eqref{eq:landau} is satisfied to numerical precision.

\bigskip
For the derivation of the DSE we now define an infinitesimal left-variation $\lvar^b_{lm}$ of such a \emph{gauge-fixed} 
configuration $U=\{U_{rs}\}$:
\begin{equation}
 \label{eq:varU}
   \lvar^b_{lm} U_{rs} \;\equiv\; t^b U_{lm} \delta_{rl}\delta_{sm} -U_{ml}t^b \delta_{rm}\delta_{sl}
    +\theta_r U_{rs} - U_{rs}\theta_s  \,.
\end{equation}
It generates a left-variation of the link variable $U_{lm}$ followed by an infinitesimal gauge transformation that 
returns the configuration to lattice Landau gauge (LLG). The anti-hermitian traceless $n\times n$ matrices 
$\{\theta_i\}$ in \eqref{eq:varU} ensure that the configuration remains in LLG. That is, the left-variation of 
the gauge condition is zero:
\begin{equation}\label{eq:varf}
  0=\lvar^b_{lm} f^a_i =\Re\Tr \left[t^a t^b U_{lm}(\delta_{il}-\delta_{im})+ \sum_j  t^a\left(\theta_i U_{ij}- 
  U_{ij}\theta_j \right)\right]\ .\nonumber
\end{equation}
and the components of $\theta_i=\sum_c t^c \theta^{c;\,b}_{i;\;lm}$ solve the linear system,
\begin{equation}
\label{eq:lineareq}
\sum_{c,j} M^{ac}_{ij}\theta^{c;\,b}_{j;\,lm}= \mathcal{U}_{lm}^{ab}(\delta_{il}-\delta_{im})\ ,
\end{equation}
where $M$ is the Faddeev-Popov (FP) matrix
\begin{equation}\label{eq:defM}
M^{ab}_{ij}:={\cal U}_{ij}^{ba}-\delta_{ij}\sum_k{\cal U}_{ik}^{ab}\ ,
\end{equation}
given in terms of the real symmetric matrix,
\begin{equation}
 \label{eq:calU}
 \mathcal{U}_{ij}^{ab}=\mathcal{U}_{ji}^{ba}:=\Re\,\Tr\left(t^a t^b U_{ij}\right)\ .
\end{equation}
The Faddeev-Popov matrix $M^{ab}_{ij}=M^{ba}_{ji}$ is symmetric under the simultaneous exchange of color and site 
indices when Eq.~\eqref{eq:landau} holds. Global gauge invariance of $V$ implies that $\sum_i 
M^{ab}_{ij}=0$, i.e., $M^{ab}_{ij}$ has $n^2-1$ generic zero-modes. The remaining $(N-1)(n^2-1)$ eigenvalues are 
positive at each local minimum of $V$, and Eq.~\eqref{eq:defM} implies that one may choose the solution 
$\theta^{a;b}_{i;lm}$ of  \eqref{eq:lineareq} orthogonal to the zero modes.

If no gauge was fixed, the integration measure would be the Wilson measure $d\mu_W=D[U] e^{S_W[U]}$. $D[U]$ is 
the product of Haar measures $dU_{ij}$ for each oriented link variable which is invariant under left- 
as well as right- group multiplication, $d(gU_{ij})=d(U_{ij}g)=dU_{ij}$ for any $g\in SU(n)$, whereas $S_W$ is 
invariant only under lattice gauge transformations $U^g_{ij}= g_i U_{ij} g^{\dagger}_j$.

To account for (minimal) Landau gauge we introduce (in a Faddeev-Popov-like manner) a density
\begin{equation}
  \label{defrho}
  \rho_\alpha(U):=\mathcal{D}_\alpha [U] e^{-\alpha \left(S_{LG}[U]-S_{LG}[\bar{U}]\right)}\quad\text{with}\quad   
  \mathcal{D}_\alpha^{-1}[U] \equiv \int \prod_i  dg_i\; e^{-\alpha \left(S_{LG}[U^g]-S_{LG}[\bar{U}]\right)}\ ,
\end{equation}
which in the limit $\alpha\to\infty$ has support only at the absolute minima $\bar{U}$ of $S_{LG}$ and whose integral 
over the 
gauge orbit 
$\int \prod_i dg_i\; \rho_\alpha(U^g)=1$. In a first attempt we set $S_{LG}[U] = V[U]$, which for $\alpha\to\infty$ 
gives support only for $\bar{U}$, the global minimum of $V$ on the gauge orbit of $U$. We will later see that for any 
of the standard lattice implementations of minimal Landau gauge (these typically find only local minima of $V$) the 
identification $S_{LG}[U] = V[U]$ accounts only for the leading contribution in the DSE.\footnote{The effective form of 
$S_{LG}[U]$ was unclear to us at the time of the conference, but it will be further discussed and specified in a 
forthcoming publication.} Anyhow in the limit $\alpha\to\infty$ it gives for $\mathcal{D}_\alpha[U]$
\begin{equation}
 \lim_{\alpha\to\infty} \mathcal{D}_\alpha[U] = \sqrt{\det M[\bar{U}]} 
 \qquad(\text{if}\ S_{LG} \equiv V)
\end{equation}
and the expectation value of a \emph{gauge-variant} quantity $\mathcal{O}$ in minimal Landau gauge (MLG) becomes
\begin{equation}
 \langle \mathcal{O}\rangle_{MLG} = \lim_{\alpha\to\infty}\frac{1}{Z_W}\int D[U]\; \rho_\alpha(U)\;\mathcal{O}[U]\, 
e^{S_W[U]}
\end{equation}
with $Z_W = \int D[U]\,\rho_\alpha(U)\, e^{S_W[U]}$. Note that in lattice perturbation theory one commonly sets 
$S_{LG}[U] = 
f^a_if^a_i$ which gives $\vert\det M[U]\vert$ for $\mathcal{D}_\alpha[U]$ (see, e.g.,\cite{Rothe:1992nt}).

The Dyson-Schwinger equation (DSE) of a link variable in minimal Landau gauge we now obtain from a left-variation 
(see Eq.~\eqref{eq:varU}) of the expectation value $\vev{U_{rs}}_{MLG}$. This expectation value must not change 
under the left-variation and so
\begin{align}\label{equal}
 \lvar_{lm}^b\left(\lim_{\alpha\to\infty} \frac{1}{Z_W}\int D[U]\; \rho_\alpha(U)\, e^{S_W[U]}\, 
U_{rs}\right) = 0\,.
\end{align}
Applying $\lvar^b_{lm}$ to each term gives the DSE in an implicit form
\begin{align}\label{DSE}
 \lim_{\alpha\to\infty} \vev{\lvar^b_{lm} U_{rs} 
  + U_{rs} \lvar^b_{lm}\left(S_W[U]+\ln\rho_{\alpha}[U]\right)}_{MLG}=0.
\end{align}
which after some algebra becomes 
\begin{equation}
 \label{eq:DSEfin}
 c_f\vev{\Tr U_{lm}}_{MLG} (\delta_{rl}\delta_{sm} - \delta_{rm}\delta_{sl})
  =\Re\sum_a \vev{K^a_{lm}\Tr t^a U_{rs}}_{MLG}+\sum_{ab}\vev{ \theta^{a;b}_{r;lm}{\cal 
U}^{ba}_{rs}-\theta^{a;b}_{s;lm}{\cal U}^{ba}_{sr}}_{MLG}\,.
\end{equation}
Here $c_f=(n^2-1)/(2n)$ is the quadratic Casimir invariant of the fundamental representation of $SU(n)$, and the 
components of $\theta_j=\sum_a t^a\theta^{a;b}_{j;lm}$ solve Eq.\,\eqref{eq:lineareq}. The (conserved) current 
$K^a_{lm}$ is of the 
form
\begin{equation}\label{eq:defK}
  K^a_{lm}:=\Sigma^a_{lm}+\Phi^{a}_{lm}\,.
\end{equation}
where $\Sigma^a_{lm}$ arises from varying the Wilson action,
\begin{equation}\label{eq:Slm}
  \Sigma^a_{lm}:=\lvar^a_{lm}S[U]=\frac{2}{g_B^2}\Re \sum_{j,k} \Tr\, t^a U_{lm} U_{mj} U_{jk} U_{kl}\ ,
\end{equation}
and $\Phi$ from the variation of the induced measure $\rho_\alpha$
\begin{align}\label{eq:defPhi}
\Phi^a_{lm}[U]:=\lim_{\alpha\rightarrow\infty} \lvar^a_{lm}\ln (\rho_\alpha[U]) = \frac{1}{2}\Re\sum_{ij,bc} 
\tilde M^{-1\,cb}_{ji}(\delta_{im}-\delta_{il})\Tr( t^b t^c \delta_{jl}-t^c t^b \delta_{jm})t^a  U_{lm}\,.
\end{align}
Here we have used that $\rho_\infty[\bar U^g]$ is stationary with respect to gauge transformations. This also implies 
that $\Phi^a_{lm}$ is transverse, $\sum_m\Phi^a_{lm}[\bar U]=0$, in fact this is true at any minimum of $V$. Also 
$\Sigma^a_{lm}$ is transverse in the sense that $\sum_m\Sigma^a_{lm} = 0$.

The longitudinal part of the lattice DSE is \emph{algebraically} satisfied by the last term of \eqref{eq:DSEfin}:   
Summing on the index "s" and using \eqref{eq:landau} and \eqref{eq:defM} we have at any minimum of $V[U]$
\begin{equation}
 \label{eq:longDSE}
  \sum_{ab,s}\left( \theta^{a;b}_{r;lm}{\cal U}^{ba}_{rs}-\theta^{a;b}_{s;lm}{\cal U}^{ba}_{sr}\right)=-\sum_{ab,s} 
   M^{ba}_{rs}\theta^{a;b}_{s;lm} =c_f (\delta_{rl} -  \delta_{rm})\Re\, \Tr U_{lm}\,.
\end{equation}

\section{Numerical verification}

To numerically verify our DSE we have Fourier-transformed Eq.\,\eqref{eq:DSEfin} to momentum space. In this way we 
take advantage of its translation invariance to maximally reduce the statistical noise of the various Monte-Carlo 
expectation values. In momentum space the DSE reads\footnote{Note, in what follows we change our notation for the 
links back to the usual, i.e., $U_{xy}\to U_{x\mu}$.}
\begin{equation}
 \label{eq:DSEmom}
 \mathcal{V}\delta_{\mu\nu} = \Sigma_{\mu\nu}(p) + \Phi_{\mu\nu}(p) + L_{\mu\nu}(p)
\end{equation}
where $\mathcal{V}=\langle V\rangle/2n$ is the (momentum-independent) expectation value of the Morse potential in 
minimal Landau gauge, while the terms on the rhs.\ are momentum-dependent. To 
verify that these dependences exactly cancel in the sum we have calculated (on several gauge-fixed ensembles) the 
term  ($n_g=n^2-1$)
\begin{align}
 L_{\mu\nu}(p) &= \frac{1}{Nn_g}\sum_{ab,xy}e^{-ip(x-y)}\vev{
   \theta^{a;b}_{y;x\mu}\mathcal{U}^{ba}_{y\nu}-\theta^{a;b}_{y+\nu;x\mu}\mathcal{U}^{ab}_{y\nu}}
\end{align}
with $\theta$ solving Eq.\,\eqref{eq:lineareq} and the transverse terms
\begin{align}
 \Sigma_{\mu\nu}(p) &=\frac{2}{g_0^2}\frac{1}{Nn_g} \sum_{a,xy} e^{-ip(x-y)} \Re\sum_a \vev{ \left(\Re\Tr\, 
t^a U_{x\mu}
     W_{x\mu}\right) \cdot \Tr t^a U_{y\nu}}\\[-3ex]
 \intertext{and}\nonumber\\[-6ex]
 \Phi_{\mu\nu}(p)   &=\frac{1}{Nn_g}\sum_{a,xy} e^{-ip(x-y)} \vev{\Phi^a_{x\mu}\cdot \Tr t^a U_{y\nu}}\,.
\end{align}
$W_{x\mu}$ is the sum of staples attached to the link $U_{x\mu}$ and $\Phi^{a}_{x\mu}$ can 
be read off from Eq.~\eqref{eq:defPhi} identifying $l=x$ and $m=x+\hat{\mu}$. For its evaluation we need 
different elements of $M^{-1}[U]$ which we estimate with the stochastic noise technique. It is remarkable, that a 
number of 8 to 32 Gaussian noise vectors for each $U$ is sufficient to provide us with a good signal for 
$\Phi^a_{x\mu}$. 

\begin{figure*}
\centering
\mbox{\includegraphics[width=0.47\linewidth]{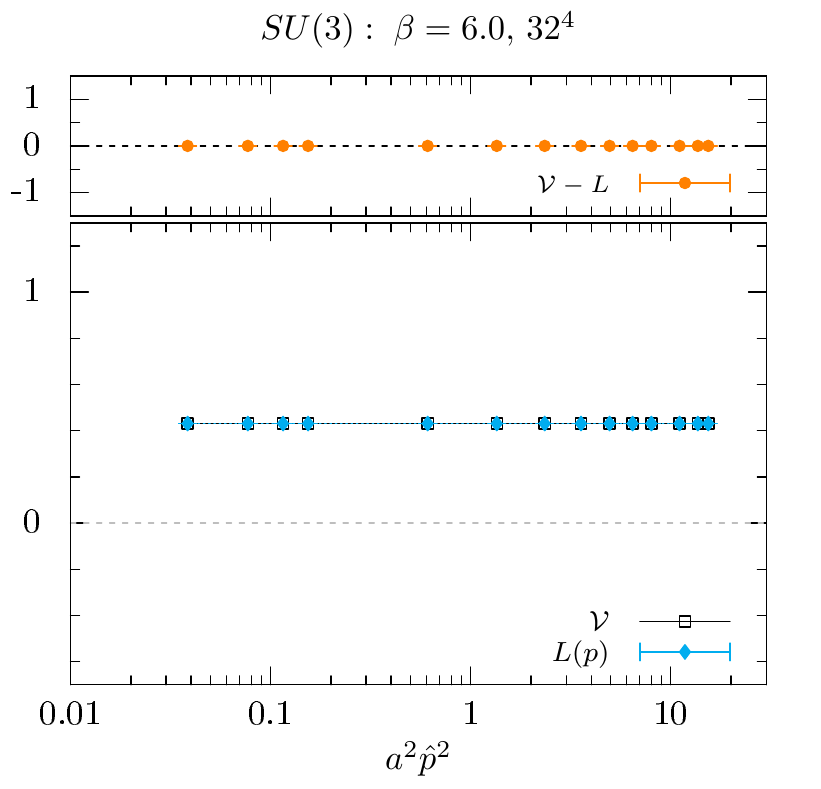}\quad
      \includegraphics[width=0.47\linewidth]{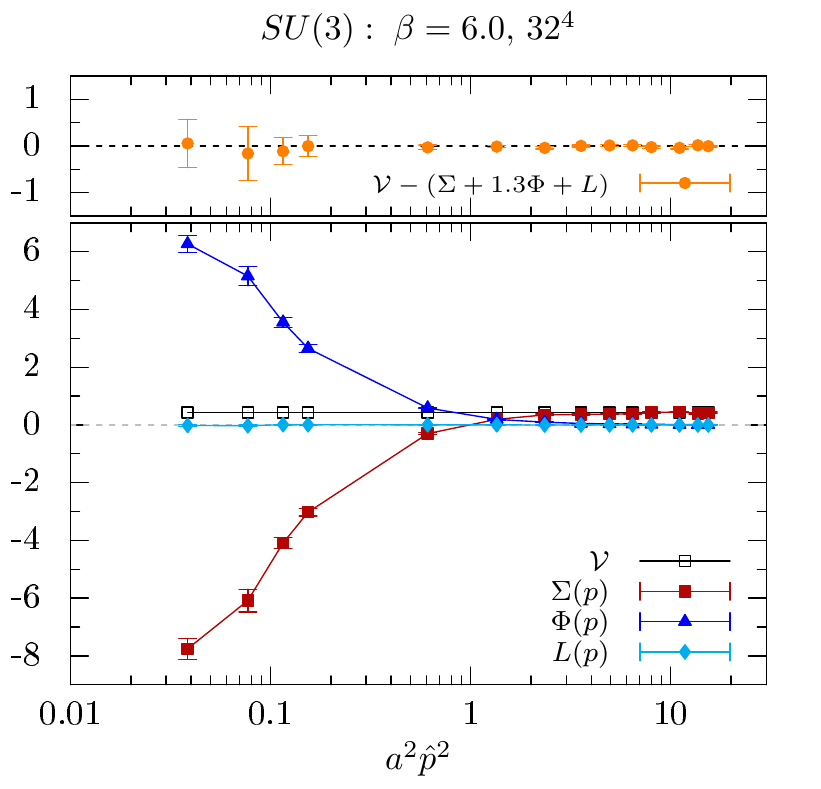}}
 \caption{Longitudinal (left) and transverse (right) terms of the DSE as a function of the lattice momentum. The 
  small panels on top show the validity of the (modified) DSE for all momenta.}
 \label{fig:DSE}
\end{figure*}

\begin{figure*}
\centering
\mbox{\includegraphics[width=0.47\linewidth]{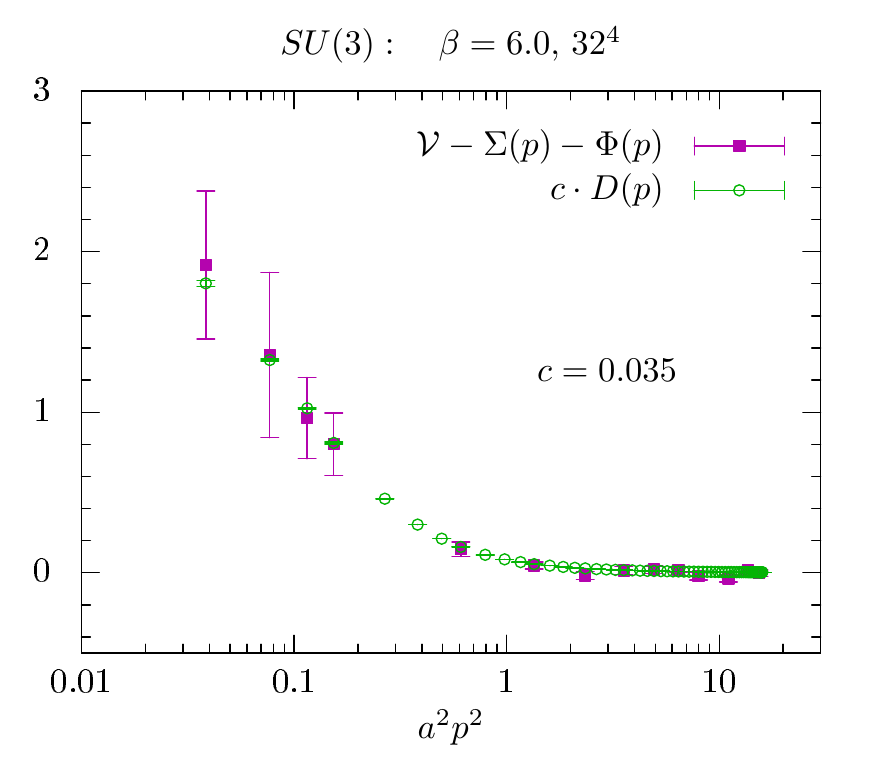}\quad
      \includegraphics[width=0.47\linewidth]{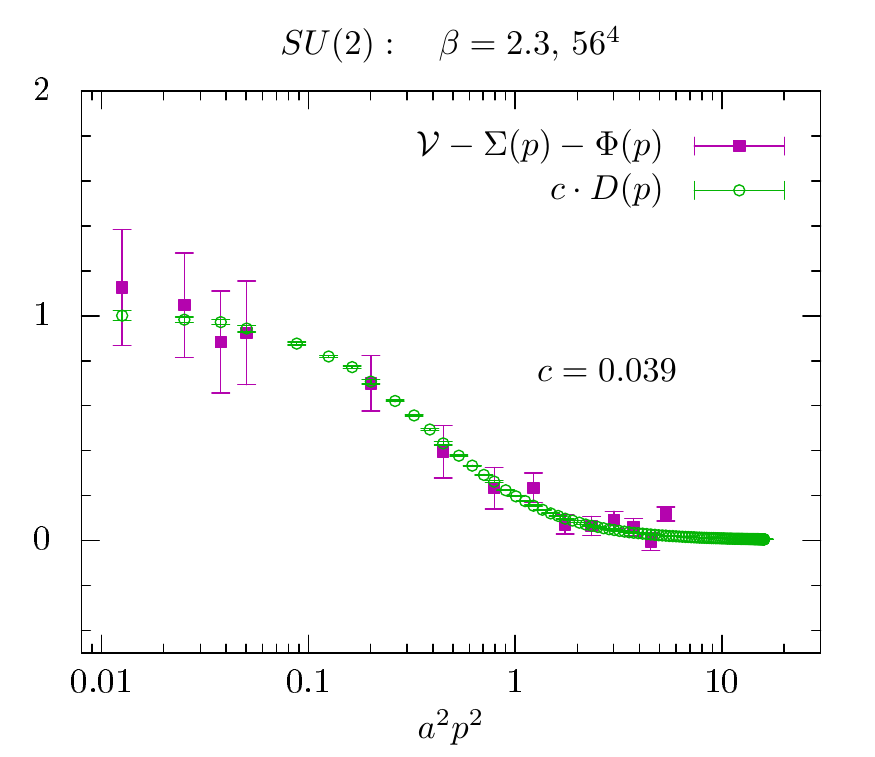}}
 \caption{Deviation of the transverse part of our initial ansatz for the DSE (full squares) in comparison to the 
corresponding gluon propagator times a constant (open diamonds). All in units of the lattice spacing and versus the 
lattice momentum squared. Left: SU(3) at $\beta=6.0, 32^4$; Right: SU(2) $\beta=2.3, 56^4$.}
 \label{fig:deviation}
\end{figure*}
In the continuum limit $L_{\mu\nu}$ becomes longitudinal. On the lattice this is not strictly the case, numerically 
however, the transverse contribution is negligible at the considered values of $\beta$.

In Fig.\ref{fig:DSE} we show the longitudinal (left panel) and transverse (right panel) terms of the DSE in momentum 
space. To this end, we have projected all terms with the respective longitudinal and transverse projectors, $P_L = 
\hat{p}_{\mu}\hat{p}_{\nu}/{\hat{p}^2}$ and $P_T = 1 - P_L$ where $a\hat{p}_\mu=2\sin(\pi k_\mu/L_\mu)$. The figure 
hence shows the ``form factors'' of $L_{\mu\nu}$, $\Sigma_{\mu\nu}$ and $\Phi_{\mu\nu}$ versus momentum 
$a^2\hat{p}^2$. This is similar to what one typically does for the gluon propagator.

In the left panel of Fig.\ref{fig:DSE} one clearly sees that the longitudinal part of the DSE is fulfilled for all 
momenta, as expected. Only $L_{\mu\nu}(p)$ gives a contribution, and the longitudinal part of 
$L_{\mu\nu}(p)-\mathcal{V}\delta_{\mu\nu}=0$ \ (see 
the small panel on top).

The transverse channel of the DSE has contributions from $\Sigma_{\mu\nu}(p)$, $\Phi_{\mu\nu}(p)$ and $L_{\mu\nu}$. 
Their 
sum however does not equal $\mathcal{V}\delta_{\mu\nu}$. In fact, we see a clear deviation from zero for 
the difference $\mathcal{V}\delta_{\mu\nu}-(\Sigma_{\mu\nu}+\Phi_{\mu\nu}+L_{\mu\nu})$ which can be compensated if one 
rescales $\Phi$ by 1.3 (see top right panel of Fig.\ref{fig:DSE}). Without rescaling the deviation is 
proportional to the gluon propagator $D_{\mu\nu}(p)$, calculated for the same lattice parameters. This we verified
for SU(2) and SU(3) for different $\beta$ and lattice sizes (see Fig.\ref{fig:deviation}). That 
is, the DSE would be fulfilled if it included another term $cD_{\mu\nu}$. At $\beta=6.0$ the proportionality constant 
$c$ is about $0.035$. Our current ansatz for the DSE thus does not fully account for the Monte Carlo data at 
small momenta.

We have not yet found a full explanation for this deviation. A subsequent analysis suggests that the reason for the 
deviation is our ansatz for lattice Landau gauge. Above we set $S_{LG}=V$ but our data at finite 
$\beta$ effectively appears to favor local minima of $S_{LG}=V+\epsilon f^2$. Such an effective  Morse potential would 
have the same minima as $V$ but a different Hessian and for small $\epsilon$ this Hessian would result in an additional 
term in the DSE proportional to the gluon propagator. At $\beta=0$, however, the deviation is not longer 
proportional to the gluon propagator. At $\beta=0$, the rescaling $\Phi\to 2\Phi$ would approximately restore the 
DSE for the considered range of momenta ($12^4$). For more details, and hopefully an explanation, we have to refer to a 
forthcoming publication.

\section{Kugo-Ojima}

Besides verifying the DSE we also want to check if the Kugo-Ojima correlator satisfies our DSE at $p=0$.
In the continuum the Kugo-Ojima correlator is given by the expectation value $\sum_b\vev{(D^y_\nu c)^b 
(D^x_\mu \bar c)^b}$. On the lattice this corresponds to $\sum_{abc}\vev{[{\cal U}^{ba}_{y\nu}c^a_y-c^a_{y+\nu}\,{\cal 
U}^{ab}_{y\nu}] [{\cal 
U}^{bc}_{x\mu}\bar c^c_x-\bar c^c_{x+\mu}{\cal U}^{cb}_{x\mu}]}$ where $\langle c^a_x \bar{c}^b_y 
\rangle_{c\bar{c}}=(M^{-1})^{ab}_{xy}$ is the Faddeev-Popov matrix. We look at this correlator again in momentum 
space where it reads
\begin{align}
  \label{eq:KOlat}
  u(p^2) = \frac{1}{Nn_g}\sum_{xy}e^{ip(x-y)}\sum_{abc}\vev{[{\cal U}^{ba}_{y\nu}c^a_y-c^a_{y+\nu}\,{\cal 
U}^{ab}_{y\nu}] [{\cal 
U}^{bc}_{x\mu}\bar c^c_x-\bar c^c_{x+\mu}{\cal U}^{cb}_{x\mu}]}\ .\nonumber
\end{align}
\begin{floatingfigure}[r]
 \parbox{6.5cm}{%
  \includegraphics[width=8.5cm]{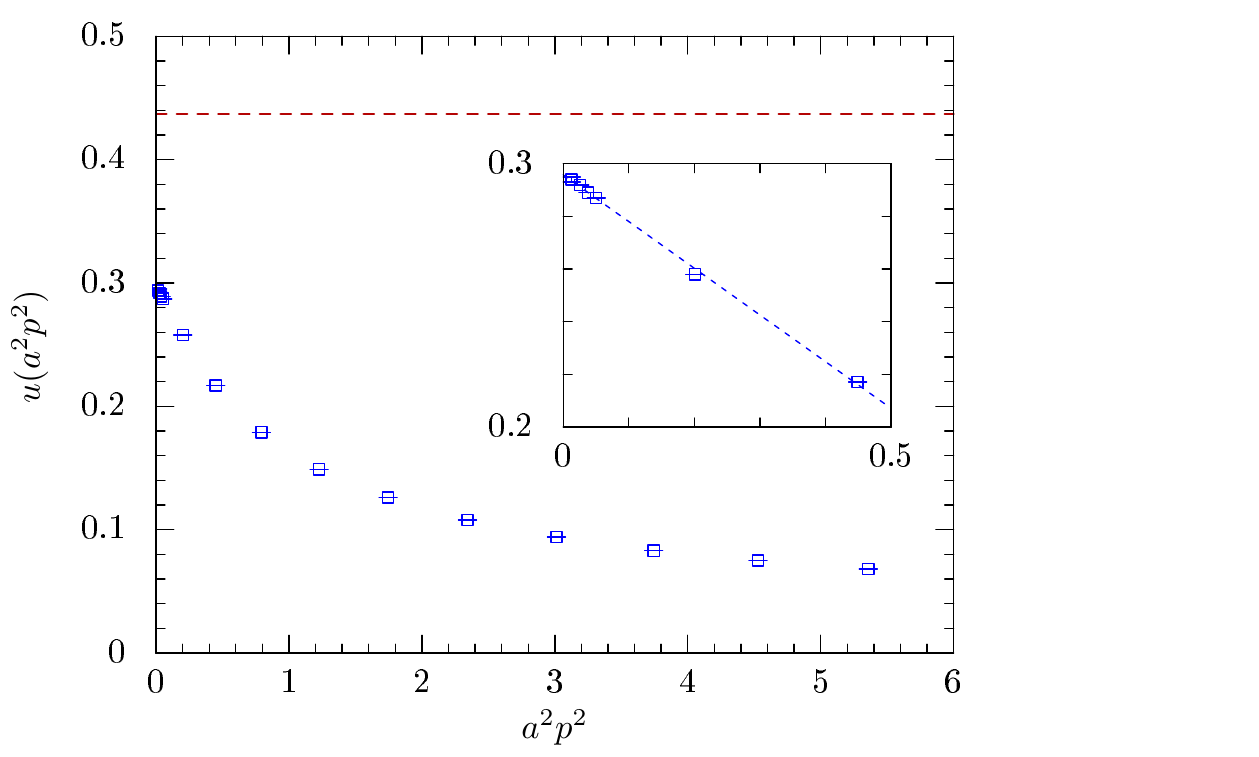}\vspace{-0.8cm}
    \caption{KO function vs.\ (lattice) momentum squared on a $56^4$ 
     lattice; $\beta=2.3$ ($n=2$). The dashed line marks the value of $\mathcal{V}$.}
     }
 \label{fig:kofunction}
\end{floatingfigure}
Since we are interested in the limit $u(p^2\to0)$, we have calculated $u(p^2)$ for the case of $SU(2)$ on a $56^4$ 
lattice at $\beta=2.3$. This has allowed us to reach relatively low momenta. This data is shown in 
Fig.~\ref{fig:kofunction}, and one clearly sees the momentum dependence behaves as expected: $u\propto p^2$ for 
$p^2\to0$. Nonetheless the limit $u(p^2\to0)$ does not equal $\mathcal{V}$, shown as dashed line in 
Fig.~\ref{fig:kofunction}. If this is a feature signaling the non-applicability of the KO criterion for 
minimal lattice Landau gauge, or related to our ansatz for the DSE needs to be clarified yet.
 
\section{Summary}

We have developed, for the first time, the Dyson-Schwinger equation for a lattice link variable in minimal Landau 
gauge. The longitudinal channel of our DSE is algebraically satisfied, but for the transverse channel we see clear 
deviations whose origin is not fully understood, but will be further analyzed in a forthcoming 
publication. Once this is settled, our data for the KO function will also be revisited again. At present it signals 
the non-applicability of the KO criterion for lattice gauge theories in Landau gauge, but this may be related to 
our present DSE.

\section*{Acknowledgements}

A.S.~acknowledges support by the European Union under the Grant Agreement IRG 256594 and the SFB/TRR-55 by the DFG. The 
data analyses was performed at the HLRN (Germany).

%

\end{document}